\documentclass[twocolumn,aps,prl,superscriptaddress,showpacs]{revtex4}
\usepackage{amssymb}
\usepackage{amsmath}
\usepackage{bm}
\usepackage{graphicx}
\usepackage{color}

\begin{document}

\title{Vanishing Meissner effect as a hallmark of in--plane FFLO instability
in superconductor -- ferromagnet layered systems}
\author{S. Mironov}
\affiliation{Institute for Physics of Microstructures, Russian Academy of Sciences,
603950 Nizhny Novgorod, GSP-105, Russia}
\author{A. Mel'nikov}
\affiliation{Institute for Physics of Microstructures, Russian Academy of Sciences,
603950 Nizhny Novgorod, GSP-105, Russia}
\author{A. Buzdin}
\affiliation{Institut Universitaire de France and University Bordeaux, LOMA UMR-CNRS
5798, F-33405 Talence Cedex, France}
\date{\today}

\begin{abstract}
We demonstrate that in a wide class of multilayered superconductor
-- ferromagnet structures (e.g., S/F, S/F/N and S/F/F$^{\prime}$)
the vanishing Meissner effect signals the appearance of the
in-plane Fulde-Ferrell-Larkin-Ovchinnikov (FFLO) modulated
superconducting phase. In contrast to the bulk superconductors the
FFLO instability in these systems can emerge at temperatures close
to the critical one and is effectively controlled by the S layer
thickness and the angle between magnetization vectors in the
F/F$^{\prime}$ bilayers. The predicted FFLO state reveals through
the critical temperature oscillations vs the perpendicular
magnetic field component.
\end{abstract}

\maketitle

The diamagnetic supercurrent and resulting magnetic field
expulsion observed in seminal experiments by Walther Meissner \cite{Meissner}
are known to be one of fundamental phenomena peculiar to
superconducting materials. The London theory \cite{London} gives us a famous
expression for the supercurrent density $\mathbf{j}=-e^2
n_s\mathbf{A}/mc$ originating from the phase rigidity of the wave
function of superconducting electrons. Here $n_s$ is the
density of superconducting electrons, $m$ is the electron
mass and $\mathbf{A}$ is the vector potential. Assuming naturally
the electron density and mass to be positive we always get the
$\mathbf{j}-\mathbf{A}$ relation corresponding to a diamagnetic
response. Recently this observation has been questioned in several
theoretical works \cite{Bergeret_Current,Asano,Yokoyama}
predicting the sign change in the London relation and an unusual
paramagnetic response of the hybrid superconductor/ferromagnet
(S/F) and superconductor/normal metal (S/N) systems. Such
anomalous Meissner effect has been attributed to the odd-frequency
spin-triplet superconducting correlations generated due to
proximity effect \cite{BergEfVolk-rew}.

For S/F systems the inversed sign of the Meissner currents
 is closely related to the oscillatory behavior of the Cooper pair wave function
inside the ferromagnet \cite{BuzRev,GolubovRMP}. These
oscillations are known to cause a number of important fingerprints
of the S/F proximity effect including
 local increase in the electronic density of states at the Fermi energy \cite{BuzdinDOS,Kontos,Braude,Cottet},
$\pi -$Josephson junction formation \cite{BuzBul,Ryazanov1} and non-monotonic dependencies of the critical
temperature of S/F bilayers on the F layer
thickness \cite{Jiang,Zdravkov}.

The unusual electromagnetic response contribution becomes even
stronger for a superconductor placed in contact with a composite
F/F$^{\prime }$ layer with different mutual orientations of the
magnetic moments. Such systems are known to reveal so--called
long-range triplet superconducting correlations predicted in Refs.~\cite{Bergeret_Current,Kadigrobov}. The local supercurrent density can be written as
${\bf j}=-e^{2}(n_{s}-n_{t}){\bf A}/mc$, where $n_{s}(n_{t})$  is the density
of the singlet (triplet) condensate. Different character of the
$n_{s}$ and $n_{t}$ components decay leads to the change in
the sign of the local response, i.e., inversion of the Meissner
effect. During the last two years an important breakthrough
in the experimental observation of the long-ranged
triplet proximity effect occurred \cite{Robinson,Khaire}. All this makes
very timely the study of the magnetic response of the proximity
induced triplet superconductivity.  Note that the first
experimental measurements \cite{Lemberger} of the London
penetration depth in thin S/F bilayers
 revealed a slightly non-monotonic dependence of
the penetration depth on the F layer thickness, which was in
accordance with the theoretical analysis \cite{HouzetMeyer}.

In this paper we address the intriguing problem of the Meissner
response of the S/F systems exhibiting the above sign change in
the relation between the supercurrent density and vector potential
and show that the anomalous Meissner effect can cause the
in--plane Fulde-Ferrell-Larkin-Ovchinnikov (FFLO) instability \cite{Fulde,Larkin}
of the superconducting uniform state. To elucidate our main
results we start from rather general arguments illustrating the
physical origin of the instability in systems with the anomalous
Meissner effect. Considering
the local supercurrent density ($\mathbf{j}=-\delta F_A/\delta \mathbf{A}%
=-e^2 n_s\mathbf{A}/mc$) as a variational derivative of the free
energy
functional we find the corresponding free energy term: $F_A=\int (e^2 n_s%
\mathbf{A}^2/2mc)dV$. The sign change in the current -- vector
potential relation can be considered as a change in the sign of
the
effective mass. Introducing the superconducting order parameter phase $%
\varphi$ and writing the free energy in the gauge--invariant form
\begin{equation}
F_A=\int\frac{e^2 n_s}{2mc}\left(\mathbf{A}-\frac{\Phi_0}{2\pi}\nabla\varphi%
\right)^2 dV,
\end{equation}
where $\Phi_0$ is the flux quantum, one can clearly see that the negative local effective mass can, in principle, result in the instability of the homogeneous state with $\varphi=const$, $\mathbf{A}=0$ and appearance of the phase
$\varphi$ modulation. Namely such a situation is realized at the
transition to the non-uniform FFLO state (see the discussion, for
example, in \cite{BuzHich}). As a consequence, the above
expression describing the linear current response should be
reconsidered for a new inhomogeneous ground state.

To illustrate the above general arguments by a concrete example of
instability we hereafter focus on the consideration of thin film
structures of total thickness much smaller than the screening
length. This assumption allows us to consider only the currents
flowing in the film plane and neglect the change of the vector
potential on the structure thickness.
Introducing the in--plane FFLO modulation vector $\mathbf{k}$ so that $%
\varphi=\mathbf{k}\mathbf{r_{\parallel}}$ we find:
\begin{equation}
F_A=\left(\mathbf{A}_{\parallel}-\frac{\Phi_0}{2\pi}\mathbf{k}\right)^2 S\int\frac{%
e^2 n_s}{2mc} dx \ ,
\end{equation}
where the $x$ axis is chosen perpendicular to the film plane, $S$
is the
sample area in the $(yz)$ plane, $\mathbf{A}_{\parallel}$ and $\mathbf{r_{\parallel}}$ are parallel to the film. All the states with $\Lambda^{-1}=\int(e^2 n_s/2mc)dx<0$ are clearly unstable and, thus, the boundary of the in--plane FFLO instability is given by the condition $\Lambda^{-1}=0$ of vanishing Meissner effect for the in--plane
field. Note that the above arguments, being applied for the FFLO
state itself, clearly show that in the modulated state the
Meissner response should be diamagnetic. Thus, in the systems
under consideration the paramagnetic Meissner response appears to
be impossible.

\begin{figure}[b!]
\includegraphics[width=0.45\textwidth]{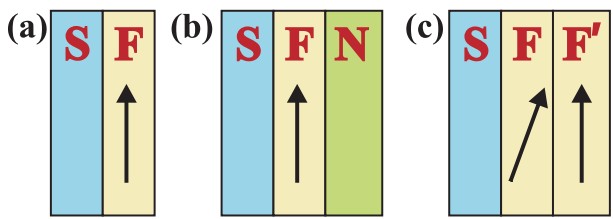}
\caption{(Color online) The sketch of the hybrid structures under
consideration. S layer is placed in contact with (a) F film, (b) F/N bilayer
and (c) F/F$^{\prime}$ bilayer with different magnetic moment orientations
shown by arrows.}
\label{Fig_1}
\end{figure}

We now proceed with the microscopic calculations of the FFLO
critical temperature and magnetic screening length for three
particular structures (S/F, S/F/N and S/F/F$^{\prime}$) shown in
Fig.~\ref{Fig_1}. Note that for S/F bilayers the modulated along
the F layer state has been suggested in Ref.~\cite{Izumov} but
later it has been pointed out \cite{FominovCrit} that the
conclusions of Ref.~\cite{Izumov} are based on the wrong boundary
conditions assuming the modulation of the order parameter only in
the F layer. In contrast with Ref.~\cite{Izumov} in our case the
same modulation is present both in S and F layers. Somewhat
similar non-uniform phase has been predicted for a ferromagnetic
cylinder covered by the superconducting shell \cite{Samokh}.
Interestingly in $^3$He films  the non-uniform superfluid p-wave
state may be  stimulated by the surface scattering of
quasiparticles \cite{sauls}.
%%%%%%%%%%%%%%%%%%%%%%%%%%%%%%%%%%%%%%%%%%%%%
%%%%%%%%%%%%%%%%%%%%%%%%%%%%%%%%%%%%%
In our calculations we assume that: (i) the system is in a dirty
limit; (ii) the exchange field $h$ in the ferromagnet is much larger than
the critical temperature $T_{c0}$ of the isolated S layer; (iii) the
thickness of the S layer $d_s$ is smaller than the coherence length $\xi_s=\sqrt{D_s/2\pi T}$ ($D_s$ is the diffusion constant in a superconductor), so
we can neglect the variation of the order parameter function $\Delta$ across
the S layer; (iv) all interfaces are transparent.

Near the critical temperature the anomalous Green function
\begin{equation}  \label{Matrix_f}
\hat{f}=\left(%
\begin{array}{cc}
f_{11} & f_{12} \\
f_{21} & f_{22} \\
\end{array}
\right)=\left(f_s+\mathbf{f}_t\hat{\sigma}\right)i\hat{\sigma}_y.
\end{equation}
satisfies the linearized Usadel equation \cite{Eschrig}
\begin{equation}  \label{Main_System}
\frac{D}{2}\nabla^2\hat{f}-\omega_n \hat{f}-\frac{i}{2}\left(\mathbf{h}
\mathbf{\hat{\sigma}} \hat{f}+ \hat{f}\mathbf{h}\mathbf{\hat{\sigma}}\right)+
\hat{\Delta}=0,
\end{equation}
where $\hat{\Delta}=\Delta i\hat{\sigma}_y$ is the superconducting
gap function, $\omega_n=\pi T(2n+1)$ are the Matsubara
frequencies, and $D$ is the diffusion constant, which may be
different for different layers. In the absence of the
barriers between layers the function $\hat{f}$ as well as the combination $\sigma\partial_x\hat{f}$ are continuous at each interface ($\sigma$ is the
Drude conductivity of the corresponding layer). We assume Fermi velocities
in all layers to be equal, so that the ratio between conductivities of
different layers is the same as the ratio between the corresponding
diffusion constants. The critical temperature $T_c$ of the system is
determined by the component $f_{12}^S$ of the Green function in the
superconductor in accordance with the self-consistency equation
\begin{equation}\label{SelfConsistencyEquation}
\Delta\ln\frac{T_c}{T_{c0}}+\sum\limits_{n=-\infty}^{\infty}
\left(\frac{\Delta}{\left|2n+1\right|}-\pi T_c f_{12}^S\right)=0,
\end{equation}
where $T_{c0}$ is the critical temperature of the isolated superconducting
film.

In the limit of weak screening the Meissner response averaged over the
structure thickness $d_0$ takes the form
\begin{equation}  \label{Lambda_def}
\lambda^{-2}=\frac{1}{\Lambda d_0}=\frac{16\pi^3 T_c}{ec\Phi_0 d_0}
\sum\limits_{n=0}^{\infty} \int\sigma\left(\left|f_s\right|^2-\left|\mathbf{f}_t\right|^2\right)dx\ .
\end{equation}
This expression clearly shows that the triplet component provides the
negative contribution to $\lambda^{-2}$. To describe the FFLO state we
assume the gap $\Delta(\mathbf{r}_\parallel)=\Delta_0\exp\left(i\mathbf{k}%
\mathbf{r}_\parallel\right)$ and the anomalous Green function $\hat{f}=\hat{%
\varphi}(x)\exp\left(i\mathbf{k}\mathbf{r}_\parallel\right) $ to be
spatially modulated. %%%%%%%%%%%%%%%%%%%%%%%%%%%%%%%%%%%%%%%%%%%%%
\begin{figure}[t!]
\includegraphics[width=0.4\textwidth]{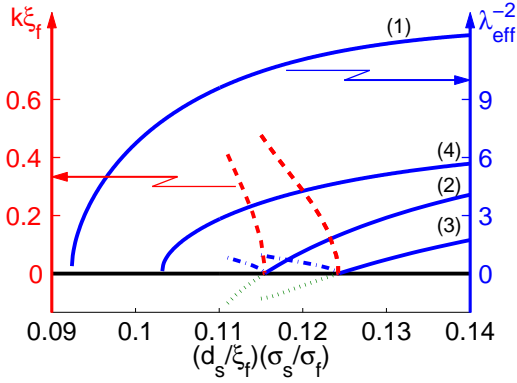}
\caption{(Color online) Magnetic screening parameter $\protect\lambda^{-2}$
(blue solid lines) in the uniform superconducting state and the optimal FFLO
modulation vector $k_0$ (red dashed curves) vs the S layer thickness $d_s$
for the S/F bilayer. Blue dash-dotted curves drawn by hand illustrate the
behavior of the magnetic screening parameter in the FFLO regime while green
dotted curves correspond to the $\protect\lambda^{-2}$ behavior
calculated for the unstable uniform state. We take $\protect\xi_{s0}=\protect%
\sqrt{D_s/4\pi T_{c0}}=0.1\protect\xi_f$ and (1) $d_f=0.75\protect\xi_f$, (2) $%
d_f=1.0\protect\xi_f$, (3) $d_f=1.2\protect\xi_f$, (4) $d_f=2.0\protect\xi_f$%
. Also we denote $\protect\lambda_{eff}^{-2}=\protect\lambda^{-2}\left(
T_c(0)ec\Phi_0 d_0/2\pi\protect\sigma_sd_s\Delta^2\right)$.}
\label{Fig_LK}
\end{figure}
%%%%%%%%%%%%%%%%%%%%%%%%%%%%%%%%%%%%%

We start from the simplest case of a bilayer (see
Fig.~\ref{Fig_1}(a)), which consists of a thin S
film and F layer of the thickness $d_f\ll \xi_n$. The
exchange field $\mathbf{h}$ in the F layer is uniform and
directed along the $z$-axis, so that $f_{11}=f_{22}=0$.
Substituting the modulated Green function into the Usadel equation
and solving it under the assumption, that the function $\hat{f}$
weakly varies across the S layer, we obtain the components
$f_{12}^{S(F)}$ in the S(F) layer
\begin{equation}  \label{SF_FFLO_Solution}
f_{12}^S=\frac{\Delta_0 e^{i\mathbf{k}\mathbf{r}_\parallel}}{\omega_n+%
\tau_{s}^{-1}(k)},~ f_{12}^F=f_{12}^S\frac{\cosh\left(q_k\left(x-d_f\right)%
\right)}{\cosh\left(q_kd_f\right)},
\end{equation}
where $q_k=\sqrt{q^2+k^2}$, $q=\left(1+i\right)/\xi_f$, and $\xi_f=\sqrt{%
D_f/h}$ is the coherence length in the ferromagnet. The complex
pair-breaking parameter
\begin{equation}  \label{tau_s}
\tau_{s}^{-1}(k)=\frac{D_s}{2}k^2+\frac{D_s}{2d_s}\frac{\sigma_f}{\sigma_s}
q_k\tanh\left(q_kd_f\right)
\end{equation}
determines the critical temperature $T_c(k)$ of the S film:
\begin{equation}  \label{SF_Tc}
\ln\frac{T_c(k)}{T_{c0}}=\Psi\left(\frac{1}{2}\right)-\mathrm{Re}\Psi\left(
\frac{1}{2}+\frac{\tau_{s}^{-1}(k)}{2\pi T_c(k)}\right) \ ,
\end{equation}
where $\Psi$ is the Digamma function. Note that these results can be
obtained by replacing $\omega_n\to\omega_n+D_{s(f)} k^2/2$ in the Usadel equation
for the uniform state.

The effective magnetic screening length in the uniform state can be
expressed through the derivative of the above expression for $T_c$ at $k=0$:
\begin{equation}  \label{Lambda_res}
\lambda^{-2}=-\frac{d_s\sigma_s\Delta^2 D_s}{2ec\Phi_0 d_0 T_c(0)} \left[1-
\mathrm{Re}\left\{\nu\Psi_1\left(\frac{1}{2}+\nu\right)\right\}\right] \left.
\frac{\partial T_c}{\partial k^2}\right|_{k=0}.
\end{equation}
Here $\nu=\tau_s^{-1}(0)/2\pi T_c(0)$ and $\Psi_1$ is the trigamma function. Calculating the derivative of the Eq.~(\ref{SF_Tc}) we
find the result obtained in \cite{HouzetMeyer}. The condition of the
stability of the uniform superconducting state, $\partial T_c/\partial k^2
(k=0)<0$, imposes a diamagnetic character of the Meissner response for the
magnetic field parallel to the plane of the layers.

\begin{figure}[t!]
\includegraphics[width=0.4\textwidth]{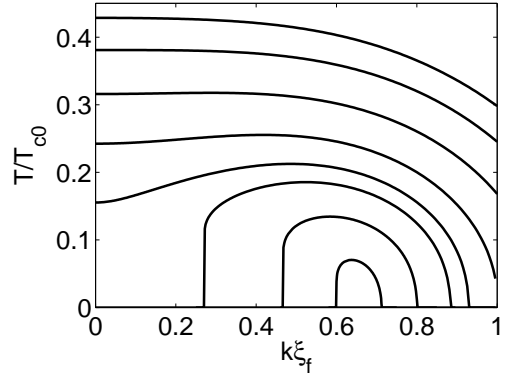}
\caption{The dependencies of the critical temperature $T_c$ vs the
modulation vector $k$ for different thicknesses $d_s$ of the S
film. We take
$\protect\xi_{s0}=\protect\sqrt{D_s/4\pi T_{c0}}=0.1\protect\xi_f$, $d_f=1.2\xi_f$ and
the
following set of values $(d_s/\protect\xi_f)(\protect\sigma_s/\protect\sigma%
_f)$: 0.13, 0.125, 0.12, 0.1165, 0.1148, 0.114, 0.113, 0.1125. The
increasing $d_s$ thickness corresponds to the increasing $T_c$
maximum.} \label{Fig_TcK}
\end{figure}

For $d_{f}\thicksim \xi _{f}$ the contribution from the F
layer to the Meissner response coefficient $\lambda ^{-2}$ can
become negative. For S/F bilayers with a large difference in the diffusion
constants ($D_{f}/D_{s}\gg h/T_{c0}$) the screening parameter $\lambda ^{-2}$
can even vanish at some critical thickness $d_{sc}\sim (\sigma _{f}/\sigma _{s})\xi _{f}$.
 At the critical thickness $d_{s}=d_{sc}$ the derivative $\left. \partial
T_{c}/\partial (k^{2})\right\vert _{k=0}$ turns to zero and $d_{s}<d_{sc}$
 the superconducting transition occurs not to the uniform
 but to the modulated FFLO state with the modulation vector $k_{0}\not=0$.
 The typical dependencies $\lambda ^{-2}(d_{s})$
 are shown by blue solid curves in Fig.~\ref{Fig_LK}. The
dependencies $k_{0}(d_{s})$ for different $d_{f}$ are shown by red
dashed curves in Fig.~\ref{Fig_LK}. The corresponding dependencies
$T_{c}(k)$ are shown in Fig.~\ref{Fig_TcK}. It is interesting that the FFLO state can survive even for parameter range
corresponding to a complete suppression of the uniform BCS state
for all temperatures.

Discussing the physical reason of the FFLO phase emerging in S/F
bilayer we should note that the proximity effect with a
ferromagnet plays a role of the pair-breaking effect at the S/F
interface. The FFLO-like modulation of the pairing wave
function weakens such pair-breaking, but at the same time this
modulation suppresses partially the critical temperature of the S
layer. The efficiency of the first (second) mechanism is governed
by the diffusion coefficient $D_{f}$($D_{s}$) and finally for
$D_{f}/D_{s}\gg h/T_{c0}$ the critical temperature of the FFLO
state may exceed that of the uniform one.

%%%%%%%%%%%%%%%%%%%%%%%%%%%%%%%%%%%%%%%%%%%%%
%%%%%%%%%%%%%%%%%%%%%%%%%%%%%%%%%%%%%
%%%%%%%%%%%%%%%%%%%%%%%%%%%%%%%%%%%%%%%%%%%%%
\begin{figure}[b!]
\includegraphics[width=0.3\textwidth]{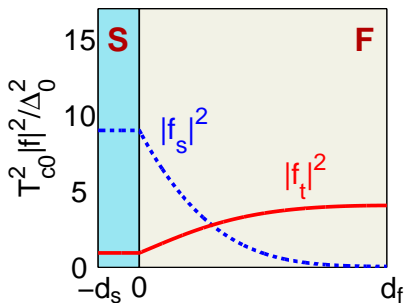}
\caption{(Color online) The spatial profile of the singlet
(blue dashed curve) and triplet (red solid curve) components
 of the anomalous Green function in the S/F bilayer
 at $T=T_c(0)$ and $\omega_n=\pi T$. We take $\xi_{s0}=0.1\xi_f$, $d_f=1.2\xi_f$
 and $\left(d_s/\xi_f\right)\left(\sigma_s/\sigma_f\right)=0.13$.}
 \label{Fig_add}
\end{figure}
%%%%%%%%%%%%%%%%%%%%%%%%%%%%%%%%%%%%%

Our analysis reveals a direct relation between the vanishing
Meissner effect and the FFLO phase formation. In
Fig.~\ref{Fig_add} we show the distribution of the triplet and
singlet components over the bilayer thickness in the BCS state
close to the threshold of the FFLO instability. One can see that
the triplet component providing the anomalous contribution to the
Meissner effect strongly exceeds the singlet one at the free
surface of the F layer. This circumstance gives a hint how to
stabilize the FFLO phase: one should add the normal metal (N)
layer on the top of the ferromagnetic layer (see Fig.~1(b)).
Moreover such modified system
allows to overcome the strong damping of $T_{c}$ in the FFLO state
of the S/F bilayer and get the FFLO state for temperatures close
to $T_{c0}$. Details of calculations can be found in Supplemental Material \cite{supp}.

The appearance of the FFLO state can be effectively controlled
provided we
consider S/F/F$^{\prime}$ structures (see Fig.1(c)) with a certain angle $%
\theta$ between the magnetization vectors in the F and F$^{\prime}$ layers.
Such systems are recently discussed as possible candidates for spin valve
devices \cite{Beasley,Tagirov,Garifullin}. For non-collinear magnetic
moments the triplet component of the anomalous Green function generated in
the F film becomes long-range in the F$^{\prime}$ layer and decays at a
distance of the order of $\xi_{n}\gg \xi_f$ (where $\xi_{n}=\sqrt{%
D_{f^{\prime}}/4\pi T_{c0}}$) while the singlet component is fully damped at
a distance $\sim \xi_{f^{\prime}}=\sqrt{D_{f^{\prime}}/h}$ from the F/F$%
^{\prime}$ interface. As a result, if the thickness $d_{f^{\prime}}$ of the F%
$^{\prime}$ layer strongly exceeds $\xi_f$ then the corresponding
contribution into the screening parameter $\lambda^{-2}$ is
\textit{always} negative and can become comparable with the one
from the S film. In the simplest case of small (large) thickness
of the
 F(F$^{\prime}$) layer, i.e. $d_f\ll \xi_f$ and $d_{f'}\to\infty$,
 the long-ranged triplet component $f_{11}^{F^{\prime}}$ in the
 F$^{\prime}$ layer is proportional to $(d_f/\xi_f)^2\sin\theta$.
  For large diffusion constant $D_{f'}$ the ratio between the
   negative contribution coming from the F$^{\prime}$ layer
    and positive S layer contribution to the screening parameter
    $\lambda^{-2}$ can become of the order of unity for
\begin{equation}
\sin^2\theta\gtrsim \frac{D_s }{D_{f\prime}} \frac{d_s}{\xi_n }
\left(\frac{\xi_f}{d_f }\right)^4 \ .
\end{equation}
Varying the angle $\theta$ one can trigger the transition from the
uniform state, realized for $\theta$ close to zero and $\pi$, to
the FFLO state, which is favorable for $\theta$ close to $\pi/2$.
 Thus, the formation of the FFLO phase should
affect the angular dependence of the critical temperature in
S/F/F' spin valves devices.

%%%%%%%%%%%%%%%%%%%%%%%%%%%%%%%%%%%%%%%%%%%%%
\begin{figure}[t!]
\includegraphics[width=0.4\textwidth]{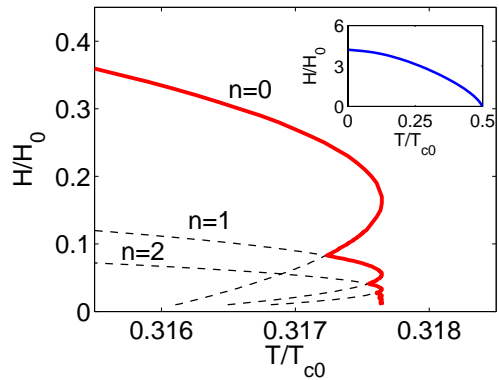}
\caption{(Color online) The phase diagram of the S/F bilayer in the FFLO
regime (red curve). Dashed curves correspond to dependencies $T_c(H)$
for different $n$. We put here $\xi_{s0}=0.1\xi_f$, $d_f=1.2\xi_f$ and $d_s=0.12(\sigma_f/\sigma_s)\xi_f$.
 For comparison in the inset we show the H-T phase diagram for $d_s=0.14(\sigma_f/\sigma_s)\xi_f$ corresponding to the
uniform superconducting state. We denote $H_0=\Phi_0/4\protect\pi \protect\xi_f^2$.}
\label{Fig_Hc2}
\end{figure}
%%%%%%%%%%%%%%%%%%%%%%%%%%%%%%%%%%%%%

Experimentally the appearance of the FFLO state can be identified by the
observation of the critical temperature oscillations vs magnetic field $H$
perpendicular to the plane of the layers \cite{Kulic}. For simplicity we consider here
only the case of a S/F bilayer. Choosing an appropriate vector potential $%
\mathbf{A}(\mathbf{r}_{\parallel})$ in the plane of the layers we get the
Usadel equation for the component $f_{12}$ in the form
\begin{equation}  \label{Usadel_H}
\frac{D}{2}\left[\partial_x^2+ \left(\partial_{\mathbf{r}_{\parallel}}-\frac{2\pi}{\Phi_0}
\mathbf{A}(\mathbf{r}_{\parallel})\right)^2\right]f_{12}-\left(\omega_n+ih%
\right) f_{12}+\Delta=0 \ .
\end{equation}
The solution of the Eq.~(\ref{Usadel_H}) takes the form: $f_{12}=\chi_n\left(%
\mathbf{r}_{\parallel}\right)\varphi(x)$, where $\chi_n\left(\mathbf{r}%
_{\parallel}\right)$ is an eigenfunction of the Hamiltonian $\hat{H}=-\left[
\partial_{\mathbf{r}_{\parallel}}-2\pi\mathbf{A}(\mathbf{r}_{\parallel})/\Phi_0\right]^2$. The critical temperature corresponding to the $n-$th
Landau level is defined by Eq.~(\ref{SF_Tc}) with $k^2\to
2\pi H(2n+1)/\Phi_0$. The competition between levels with different $n$
results in a peculiar dependence $T_c(H)$ shown in Fig.~4.

In conclusion, we predict that vanishing Meissner effect in
thin--film multilayered S/F systems should result in the in-plane
FFLO instability which is particularly important for designing the
$\pi -$junction or spin valve devices. Interestingly that in
contrast with the original FFLO phase \cite{Fulde,Larkin} which
emerges at relatively low temperature, the modulated phase in
S/F/N or S/F/F' heterostructures may appear near the critical
temperature of S layers.
 The appearance of the FFLO phase besides the anomalous behavior of the screening,
  should also result in the oscillatory-like temperature dependence of the perpendicular critical field.
   For the S layer thickness exceeding the
screening length the FFLO instability can not, of course, expand
into the bulk superconductor. In this case the surface instability
can reveal through the formation of the vortex sheet parallel (and
positioned close) to the S/F interface suppressing, thus, the
anomalous part of the Meissner response. The linear vorticity
density should be proportional to the FFLO modulation vector
$k_{0}$. Note in conclusion that similar instabilities could also
appear for the superconductors with anisotropic pairing where the
paramagnetic currents are caused by the surface -- induced Andreev
bound states \cite{Walter,Zare}.

The authors thank A.\ Samokhvalov and V.\ V.\ Kurin for useful
discussions. This work was supported by the European IRSES program
SIMTECH, French ANR ``MASH", program of LEA Physique Theorique et
Matiere Condensee, the Russian Foundation for Basic Research, the
``Dynasty'' Foundation and FTP ``Scientific and educational
personnel of innovative Russia in 2009--2013".

\end{document}